\documentclass[prb,preprint]{revtex4}

\usepackage{amsmath}

\begin{document}
\newcommand{\bmu}{\boldsymbol{\mu}}
\newcommand{\rv}{\boldsymbol{r}}
\newcommand{\ruv}{\boldsymbol{\hat r}}
\newcommand{\pv}{\boldsymbol{p}}
\newcommand{\puv}{\boldsymbol{\hat p}}
\newcommand{\grad}{\boldsymbol{\nabla}}

\title{Comment on ``Some novel delta-function identities"}

\author{Jerrold Franklin}
\email{Jerry.F@temple.edu}
\affiliation{Department of Physics, Temple University, Philadelphia, Pennsylvania 19122-6082}
\date{\today}

\begin{abstract}
We show that a form for the second partial derivative of $1/r$ proposed by Frahm\cite{f} and subsequently used by other workers applies only when averaged over smooth functions.
We use dyadic notation to derive a more general form without that restriction.
\end{abstract}

\maketitle

In Ref.~\onlinecite{f} the equation
\begin{equation}
\partial_i\partial_j\left(\frac{1}{r}\right)=
-\left(\frac{4\pi}{3}\right)\delta_{ij}\delta(\rv)+\frac{3x_i x_j-r^2\delta_{ij}}{r^5}
\label{eq:f}
\end{equation}
is proposed for the second partial derivative of $1/r$. This result has subsequently been used in
more recent papers
\cite{bowen, est, leung, zus}. The purpose of this comment is to show that the derivation of Eq.~\eqref{eq:f} in Ref.~1 is flawed, and to present a direct derivation of this second partial derivative. While Ref.~1 uses two indirect methods to deduce Eq.~(\ref{eq:f}), we employ dyadic notation\cite{jf1} to take the partial derivatives directly.

In dyadic notation, the left-hand side of Eq.~(\ref{eq:f}) can be operated on as
\begin{subequations}
\begin{align}
\grad\grad\left(\frac{1}{r}\right)&=-\grad\left(\frac{\rv}{r^3}\right)\\
&= -\frac{\grad \rv}{r^3}-\rv\grad\left(\frac{1}{r^3}\right)\\
&= -\frac{\bf{\hat{\hat I}}}{r^3}-\rv\grad\left(\frac{1}{r^3}\right),
\label{eq:dyad}
\end{align}
\end{subequations}
where ${\bf\hat{\hat {I}}}$ is the unit dyadic.

To evaluate the term $\grad(1/r^3)$,
we start with the well known identity
\begin{equation}
\grad\cdot\left(\frac{\rv}{r^3}\right) =4\pi\delta(\rv).
\label{eq:delta}
\end{equation}
The left-hand side of Eq.~\eqref{eq:delta} contains no vector other than $\rv$. Therefore the scalar functions on each side of the equation can have no angular dependence. The identity of Eq.~({\ref{eq:delta}) is usually proven by integrating the left-hand side over a volume, and applying the divergence theorem. The volume integral over the left-hand side becomes an integral over the bounding surface. Since the integrand has no angular dependence,
the integral over the solid angle equals $4\pi$ if the volume contains the origin. It equals zero if the volume does not contain the origin. Thus, the right-hand side of 
Eq.~({\ref{eq:delta}) equals $4\pi\delta(\rv)$ by the definition of the delta function, and the identity is proven.

We can express the left-hand side of Eq.~(\ref{eq:delta}) as
\begin{equation}
\grad\cdot\left(\frac{\rv}{r^3}\right)
=\frac{\grad\cdot\rv}{r^3}+\rv\cdot \grad\left(\frac{1}{r^3}\right)
=\frac{3}{r^3}+\rv\cdot \grad\left(\frac{1}{r^3}\right),
\label{eq:deriv}
\end{equation}
which isolates the term $\grad(1/r^3)$. Since the function $(1/r^3)$ depends only on $r$, its gradient must be in the $\ruv$ direction. Thus we can write
\begin{equation}
\grad\left(\frac{1}{r^3}\right) ={\ruv}g(r),
\label{eq:g}
\end{equation}
where the scalar function $g(r)$ is give by
\begin{equation}
g(r)=\ruv \cdot \grad\left(\frac{1}{r^3}\right) .
\label{eq:gr}
\end{equation}
We combine Eq.~\eqref{eq:gr} with Eqs.~({\ref{eq:delta}) and (\ref{eq:deriv}) to find
\begin{equation}
g(r)=\frac{4\pi\delta(\rv)}{r}-\frac{3}{r^4},
\label{eq:gd}
\end{equation}
and then
\begin{equation}
\grad\left(\frac{1}{r^3}\right)
=\frac{4\pi{ \ruv}\delta(\rv)}{r}-\frac{3\ruv}{r^4}.
\label{eq:gdr}
\end{equation}
Substituting Eq.~\eqref{eq:gdr} into Eq.~(\ref{eq:dyad}), we obtain
\begin{equation}
\grad \grad\left(\frac{1}{r}\right)=
\frac{3\ruv\ruv}{r^3}-\frac{{\bf \hat{\hat I}}}{r^3}
-4\pi\ruv\ruv\delta(\rv).
\label{eq:end}
\end{equation}
In the Cartesian tensor notation of Ref.~1, Eq.~\eqref{eq:end} would be written as
\begin{equation}
\partial_i\partial_j\left(\frac{1}{r}\right)=
\frac{3x_i x_j-r^2\delta_{ij}}{r^5}-\frac{4\pi x_i x_j\delta(\rv)}{r^2},
\label{eq:fc}
\end{equation}
which differs from Eq.~(\ref{eq:f}) in the delta function term.

Objections have been raised about the relevance of $x_i/r$ multiplying a delta 
function,\cite{bowen} because $x_i/r$ is not well defined in the limit $r\rightarrow 0$. However the same objection could be raised against the delta function itself, which is also undefined at the origin.  As with the delta function, factors of $x_i/r$ give a definite result when used in a volume integral, even when multiplied by a delta function. Also, our Eq.~(\ref{eq:gdr}) shows that the gradient of $1/r^3$ could not be written without $\hat \rv$ (with Cartesian component $x_i/r$) multiplying the delta function. The ratio $x_i/r$ is used in several places in Ref.~1 for infinitesimal $r$. It is inconsistent to preclude $x_i/r$ in one place and then use it in another.

Where does Ref.~1 lose the $x_i/r$ factors multiplying the delta function?
Reference 1 arrives at its expression for $\partial_i\partial_j(1/r)$ by first using a ``plausibility argument" and then a ``physicist's proof." The plausibility argument depends on the statement: ``Noting that the mixed second derivatives cannot contain a delta function\ldots ." The falsity of this statement is demonstrated by our result in Eq.~(\ref{eq:fc}) which shows that the mixed second derivative does contain a delta function, and 
does not depend on how the first derivative was taken. The plausibility argument arrives at the form $\delta_{ij}$ instead of $x_i x_j/r^2$ by incorrectly assuming it.

The physicist's proof in Ref.~1 uses an integral over solid angle to deduce that the form in Eq.~(\ref{eq:f}) is correct. This proof uses the identity
\begin{equation}
\oint\!d\Omega\,\frac{x_i x_j}{R^2}=\frac{4\pi}{3}\delta_{ij}
\label{eq:do}
\end{equation}
for the integral over the surface of a sphere of infinitesimal radius $R$.
In dyadic notation, this identity is
\begin{equation}
\oint\ruv\ruv\,d\Omega=\frac{4\pi}{3}{{\bf \hat{\hat I}}}.
\label{eq:dy}
\end{equation}
Reference~1 then uses Eq.~(\ref{eq:do}) to show that its form for the delta function term gives a correct integral when multiplied by ``an arbitrary smooth function" and integrated over a sphere with infinitesimal radius. This derivation works because a function that is smooth at the origin has a Taylor expansion which has no angular dependence in the limit of vanishing radius,
and because Eq.~(\ref{eq:do}) shows that our form for the delta function term reduces to that of Ref.~1 when averaged over solid angle. However this averaging procedure restricts the applicability of the form in Ref.~1 to smooth functions. For instance, a function such as 
$f(\puv,\ruv)=(\puv \cdot\ruv)(\puv \cdot \ruv)$,
which is not smooth at the origin, gives a different result in a volume integral when
multiplied by ${\bf \hat{\hat I}}\delta(r)$ instead of $\ruv\ruv\delta(\rv)$.
That is,
\begin{equation}
\int (\puv\cdot\ruv)(\puv \cdot\ruv)
\ruv\ruv\delta(\rv)dV=\frac{4\pi}{15} {\bf\hat{\hat{I}}}
+\frac{8\pi}{15} \puv\puv,
\label{eq:int1}
\end{equation}
while
\begin{equation}
\int (\puv \cdot\ruv)( \puv \cdot\ruv)
{\bf \hat{\hat I}}\,\delta(\rv)dV=\frac{4\pi}{3}{\bf \hat{\hat I}}.
\label{eq:int2}
\end{equation}

Even though it is incorrect, the use of Eq.~(1) in most physics applications leads to the correct result, since functions used in physics are usually smooth at the origin.
For instance, most electromagnetism textbooks derive the singular part of the electric field of an electric dipole $\pv$ by applying the divergence theorem, and effectively averaging $\bf E$ over all solid angle. In this case, either equation (\ref{eq:f}) or (\ref{eq:end}) gives the same result 
$(-4\pi \pv/3)\delta(\rv)$ for the singular part of $\bf E$.
However, writing
\begin{equation}
{\bf E}=\frac{3 (\pv \cdot\ruv)\ruv-\pv}{r^3}-\frac{4\pi\pv}{3}\delta(\rv)
\label{eq:dip}
\end{equation}
is inconsistent mathematically, because the delta function term is averaged over solid angle, while the first term is not. The use of Eq.~(\ref{eq:end}) leads to the mathematically consistent equation\cite{jf2}
\begin{equation}
{\bf E}=\frac{3(\pv \cdot\ruv)\ruv-\pv}{r^3}
-4\pi(\pv \cdot\ruv)\rv\delta(\rv).
\label{eq:myp}
\end{equation}

The magnetic field of a magnetic dipole $\bmu$ also involves the gradient of $1/r^3$ as can be seen by writing\cite{jf3}
\begin{subequations}
\begin{align}
{\bf B}&={\bf \grad\times A}
=\grad\times\left(\frac{\bmu{\bf \times r}}{r^3}\right) \\
&= \bmu \grad\cdot\left(\frac{\rv}{r^3}\right)-\bmu\cdot\grad\left(\frac{\rv}{r^3}\right)\\
&=\bmu \grad\cdot\left(\frac{\rv}{r^3}\right)-\frac{\bmu}{r^3}
-\rv\bmu\cdot \grad\left(\frac{1}{r^3}\right)\\
&=\frac{3(\bmu\cdot\ruv)\ruv-\bmu}{r^3}
-4\pi(\bmu\cdot\ruv)\ruv\delta(\rv)+4\pi\bmu\delta(\rv).
\label{eq:bmu}
\end{align}
\end{subequations}
For the last step we have used Eq.~(\ref{eq:delta}) for the divergence of $\rv/r^3$, and Eq.~(\ref{eq:gdr}) for the gradient of $1/r^3$.
We see that the field of a magnetic dipole is like that of an electric dipole, but
has an additional singular term $4\pi\bmu\delta(\rv)$. The singular part of Eq.~(\ref{eq:bmu}),
averaged over all solid angle, is given by $(+8\pi{\bmu}/3)\delta(\rv)$, which is the form
given in most textbooks for the singular part of the magnetic field of a magnetic dipole.

In summary, we have shown that the second partial derivative of $1/r$ can be found by direct differentiation using dyadic notation. As a mathematical statement, we have shown that Eq.~(1), as proposed by Ref.~1, cannot be used with functions that are not smooth at the origin. The higher derivatives considered in Ref.~1 and Refs.~\onlinecite{bowen, est, leung, zus} would also be affected by using Eq.~(\ref{eq:end}) rather than Eq.~(1). (We do not consider higher derivatives here
to keep the paper relatively simple and accessible.)

\end{document}